\documentclass[showpacs,prc,preprint,nofootinbib,showkeys]{revtex4}
\pdfoutput=1

\usepackage{bm}
\usepackage{amsmath}
\usepackage{graphicx}
\usepackage{subfigure}
\usepackage[colorlinks=true,linktocpage=true,linkcolor=blue,citecolor=blue]{hyperref}
\usepackage[usenames]{color}

\begin{document}

\title{Exact solution of the (0+1)-dimensional Boltzmann equation for a massive gas}

\author{Wojciech Florkowski}
\affiliation{The H. Niewodnicza\'nski Institute of Nuclear Physics, Polish Academy of Sciences,
PL-31342 Krak\'ow, Poland} 
\affiliation{Institute of Physics, Jan Kochanowski University, PL-25406~Kielce, Poland}

\author{Ewa Maksymiuk}
\affiliation{Institute of Physics, Jan Kochanowski University, PL-25406~Kielce, Poland}

\author{Radoslaw Ryblewski}
\affiliation{Department of Physics, Kent State University, Kent, OH 44242 United States}
\affiliation{The H. Niewodnicza\'nski Institute of Nuclear Physics, Polish Academy of Sciences, PL-31342 Krak\'ow, Poland} 

\author{Michael Strickland} 
\affiliation{Department of Physics, Kent State University, Kent, OH 44242 United States}

\begin{abstract}
We solve the one-dimensional boost-invariant kinetic equation for a relativistic massive 
system with the collision term treated in the relaxation time approximation.  The result 
is an exact integral equation which can be solved numerically by the method of iteration 
to arbitrary precision.  We compare predictions for the shear and bulk viscosities of a 
massive system with those obtained from the exact solution.  Finally, we compare the time 
evolution of the bulk pressure obtained from our exact solution with results obtained from 
the dynamical equations of second-order viscous hydrodynamics.
\end{abstract}


\maketitle 

\section{Introduction}
\label{sect:intro}

In order to properly understand the data generated in relativistic heavy-ion collisions it is necessary 
to have dynamical models that can accurately describe the time evolution of the system from the moment
after the Lorentz-contracted nuclei pass through one another, to the final production of the hadrons
that are detected.  To date, the primary tool used for describing the time evolution of the matter
created in heavy-ion collisions has been relativistic viscous  
hydrodynamics~\cite{Israel:1976tn,Israel:1979wp,Muronga:2001zk,Muronga:2003ta,Heinz:2005bw,
Baier:2006um,Baier:2007ix,Romatschke:2007mq,Dusling:2007gi,Luzum:2008cw,Song:2008hj,El:2009vj,
PeraltaRamos:2010je,Denicol:2010tr,Denicol:2010xn,Schenke:2010rr,Schenke:2011tv,Bozek:2009dw,
Bozek:2011wa,Niemi:2011ix,Niemi:2012ry,Bozek:2012qs,Denicol:2012cn,PeraltaRamos:2012xk,Jaiswal:2013npa}.  
Originally, most practitioners relied on the Israel-Stewart framework for obtaining the necessary
viscous hydrodynamic equations; however, recently there have been efforts to provide more complete
formulations of second- and third-order viscous hydrodynamics which should, in principle, more accurately
describe the time evolution of the system.  In addition to these developments, recently a framework called 
dissipative anisotropic hydrodynamics has been developed which attempts to improve upon standard relativistic 
viscous hydrodynamics approximations by relaxing the assumption that the system is approximately 
isotropic in momentum-space~\cite{Florkowski:2010cf,Martinez:2010sc,Ryblewski:2010bs,
Martinez:2010sd,Ryblewski:2011aq,Florkowski:2011jg,Martinez:2012tu,Ryblewski:2012rr,Florkowski:2012as,
Florkowski:2013uqa,Bazow:2013ifa,Tinti:2013vba,Florkowski:2014txa}.  

If one wants to assess how well these various dissipative relativistic hydrodynamics approaches 
describe the true non-equilibrium evolution of the system, it is necessary to have some exactly 
solvable cases that can be used to discriminate the various approaches.  One possible avenue for
doing this is to compare predictions of hydrodynamic models with exact solutions
of the underlying kinetic theory.  Doing this in general is not possible, however, there are
some cases in which this can be done.  Recently it was shown that it was possible to exactly
solve the Boltzmann equation in the relaxation time approximation for a system of massless 
particles which is transversely homogeneous and boost invariant~\cite{Florkowski:2013lza,
Florkowski:2013lya}.  In this paper we generalize the results obtained in Refs.~\cite{Florkowski:2013lza,
Florkowski:2013lya} to a system of massive particles.  This generalization allows us to directly 
test various predictions available in literature for the massive near-equilibrium transport 
coefficients now including bulk viscous effects.

The structure of the paper is as follows: In Section~\ref{sect:kineq} we present the kinetic
equation we solve, list the thermodynamic functions for an equilibrium massive Boltzmann gas, specify the 
boost-invariant variables we will use for the exact solution, and discuss the constraint implied by energy-momentum conservation.  In Section~\ref{sect:formsol}
we present our exact solution and associated quantities.  In Section~\ref{sect:shearbulk} we
collect results for the shear and bulk viscosities of a massive Boltzmann gas and compute some
asymptotic limits of these transport coefficients.  In this section we also list three
evolution equations for the bulk pressure which can be found in the
literature.  In Section~\ref{sect:results} we present numerical evaluation of our exact solution
and compare the results obtained with the massless limit and various results available from 
relativistic viscous
hydrodynamics.  In Section~\ref{sect:taeqT} we briefly discuss the implications of having a
fixed shear viscosity to entropy density ratio on the equilibration of the system.  Finally,
in Section~\ref{sect:concl} we conclude and give an outlook for the future.

\section{The Boltzmann equation in relaxation time approximation}
\label{sect:kineq}

In this paper we consider the relativistic Boltzmann equation 
\begin{equation}
 p^\mu \partial_\mu  f(x,p) =  C[f(x,p)] \, ,
\label{kineq}
\end{equation}
where $f(x,p)$ is the one-particle distribution function, and $C$ is the collision term which 
we treat in the relaxation time approximation~\cite{1954PhRv...94..511B},
\begin{equation}
C[f] = - \frac{p \cdot u}{\tau_{\rm eq}} (f - f_{\rm eq}) \, ,
\label{col-term}
\end{equation}
where $p \cdot u \equiv p_\mu u^\mu$ and $\tau_{\rm eq}$ is the relaxation time. Herein, we will take the background 
equilibrium distribution function $f_{\rm eq}$ to be a classical Boltzmann distribution
\begin{equation}
f_{\rm eq} = \frac{2}{(2\pi)^3} \exp\left(- \frac{p \cdot u}{T} \right) ,
\label{Boltzmann}
\end{equation}
however, we note that the results contained herein can be straightforwardly generalized to the case of Bose-Einstein or Fermi-Dirac distributions.  The factor of two in Eq.~(\ref{Boltzmann}) accounts for spin degeneracy. The temperature $T$ is obtained from the Landau matching condition which demands that the energy density calculated from the distribution function $f$ is equal to the energy density obtained from the equilibrium distribution $f_{\rm eq}$.  We will provide the details of how this is accomplished in practice below.  If the system is close to thermal equilibrium, then $T$ can be interpreted as the true temperature of the system; however, since we consider a non-equilibrium system, $T$ should be interpreted as an effective temperature which is related to the non-equilibrium energy density of the system. The quantity $u^\mu$ in Eq.~(\ref{Boltzmann}) is the flow velocity of matter with $u^\mu_{\rm LRF} = (1,{\bf 0})$ in the local rest frame (LRF) of the matter.

We note that the simple forms of Eqs.~(\ref{kineq})-(\ref{Boltzmann}) used herein are motivated in large part by the fact that there are many results which have been obtained with these assumptions and, as a consequence, this allows us to make direct comparisons with other approaches. In particular, we note that there exist several calculations of the relaxation time approximation kinetic coefficients using this setup, see e.g. Refs.~\cite{Bozek:2009dw,Anderson:1974,Czyz:1986mr,Dyrek:1986vv,cerc,Sasaki:2008fg,Romatschke:2011qp}. 

\subsection{Equilibrium thermodynamic functions}
\label{sect:eqthermo}

For massive particles obeying classical Boltzmann statistics the equilibrium particle density, entropy 
density, energy density, and pressure can be expressed as~\cite{GLW,florkowski2010}
\begin{subequations} 
\begin{align}
{\cal N}_{\rm eq}(T) &= \frac{g_0 m^2 T}{\pi^2} K_2\left( \frac{m}{T}\right), \label{neq} \\
{\cal S}_{\rm eq}(T) &= \frac{g_0 m^2}{\pi^2} 
 \left[ 4T K_{2}\left( \frac{m}{T} \right) +mK_{1} \left( \frac{m}{T}\right) \right],
\label{sigmaeq} \\
{\cal E}_{\rm eq}(T) &= \frac{g_0 m^2 T}{\pi^2} 
 \left[ 3T K_{2}\left( \frac{m}{T} \right) +mK_{1} \left( \frac{m}{T}\right) \right], 
\label{epsiloneq} \\
{\cal P}_{\rm eq}(T) &= \frac{g_0 m^2 T^2}{\pi^2} K_2\left( \frac{m}{T}\right),
\label{Peq}
\end{align}
\end{subequations}
where $K_n$ are modified Bessel functions and $g_0$ is a degeneracy factor which accounts for all internal degrees of freedom except the spin, which we have included separately in Eq.~(\ref{Boltzmann}).

\subsection{Boost-invariant variables}
\label{sect:boostinvvar}

As mentioned previously, in this paper we consider the case of a transversely homogeneous boost-invariant system. For one-dimensional boost-invariant expansion, all scalar functions of space and time can depend only on the proper time $\tau = \sqrt{t^2-z^2}$. In addition, in this case the hydrodynamic flow $u^\mu$ should have the Bjorken form in the lab frame $u^\mu = \left(t/\tau,0,0,z/\tau\right)$~\cite{Bjorken:1982qr}.  As usual, the phase-space distribution function $f(x,p)$ itself transforms as a scalar under Lorentz transformations.  In this case, the requirement of boost invariance implies that $f(x,p)$ can depend only on $\tau$, $w$, and $\vec{p}_T$ with~\cite{Bialas:1984wv,Bialas:1987en}
\begin{equation}
w =  tp_L - z E \, .
\label{w}
\end{equation}
Using $w$ and $p_L$ one can define another boost-invariant variable
\begin{equation}
v(\tau,w,p_T) = Et-p_L z = 
\sqrt{w^2+\left( m^2+\vec{p}_T^{\,\,2}\right) \tau^2} \, .  
\label{v}
\end{equation}
From Eqs.~(\ref{w}) and (\ref{v}) one finds the energy and the longitudinal momentum of a particle
\begin{equation}   
E= p^0 = \frac{vt+wz}{\tau^2},\quad p_L=\frac{wt+vz}{\tau^2} \, .  
\label{p0p3}
\end{equation}
The momentum-space integration measure can be expressed in terms of these variables as
\begin{equation}
dP = 2 \, d^4p \, \delta \left( p^2-m^2\right) \theta (p^0)
= d^2p_T \frac{dp_L}{p^0} = d^2p_T\frac{dw}{v} \, .  
\label{dP}
\end{equation}

Using the boost-invariant variables introduced above, the kinetic equation may be written in the simple form
\begin{equation}
\frac{\partial f}{\partial \tau} = 
\frac{f_{\rm eq}-f}{\tau_{\rm eq}},
\end{equation}
where the boost-invariant form of the equilibrium distribution function  (\ref{Boltzmann}) is
\begin{equation}
f_{\rm eq}(\tau, w, p_T) =
\frac{2}{(2\pi)^3} \exp\left[
- \frac{\sqrt{w^2+ \left( m^2+p_T^2 \right) \tau^2}}{T(\tau) \tau} \,  \right]. 
\label{Geq}
\end{equation}
Below, we assume that $f(\tau,w,\vec{p}_T)$ 
is an even function of $w$ and depends only on the magnitude of the transverse momentum $\vec{p}_T$, namely
\begin{equation}
f(\tau,w,p_T) = f(\tau,-w,p_T) \, .
\label{symofg}
\end{equation}

\subsection{Energy-momentum conservation}
\label{sect:enmomcom}

The energy-momentum tensor can be obtained via
\begin{equation}
T^{\mu\nu}(\tau) = g_0 \int dP \, p^\mu p^\nu f(\tau,w,p_T) \, . 
\label{Tmunu1}
\end{equation}
Using Eq.~(\ref{symofg}) one can express the energy momentum tensor (\ref{Tmunu1}) in the form \cite{Florkowski:2011jg,Martinez:2012tu}
\begin{equation}
T^{\mu\nu} = ({\cal E} + {\cal P}_T) u^\mu u^\nu - {\cal P}_T g^{\mu\nu} + ({\cal P}_L-{\cal P}_T) z^\mu z^\nu 
\, ,
\label{Tmunu2}
\end{equation}
where the energy density, ${\cal E}$, the longitudinal pressure, ${\cal P}_L$, and the transverse pressure, ${\cal P}_T$, can be obtained via
\begin{subequations} 
\begin{align}
{\cal E}(\tau) &= \frac{g_0}{\tau^2}\,
\int dP \, v^2\,  f(\tau,w,p_T) \, , \\
{\cal P}_L(\tau) &= \frac{g_0}{\tau^2}\,
\int dP \, w^2\,  f(\tau,w,p_T) \, ,  \\
{\cal P}_T(\tau) &= \frac{g_0}{2}\,
\int dP \, p_T^2\, f(\tau,w,p_T) \, ,
\end{align}
\label{epsandpres}
\end{subequations}
and $z^\mu = \left(z/\tau,0,0,t/\tau\right)$ is a four-vector which defines the beam direction. 
Energy-momentum conservation requires that
\begin{equation}
\partial _\mu T^{\mu \nu }=0 \, .
\label{enmomcon1}
\end{equation}
For a one-dimensional boost-invariant system, the four equations implicit in Eq.~(\ref{enmomcon1}) reduce to a single equation
\begin{equation}
\frac{d{\cal E}}{d\tau}=
- \frac{{\cal E}+{\cal P}_L}{\tau} \, .
\label{enmomcon2}
\end{equation}
We note that the structure of the energy-momentum tensor (\ref{Tmunu2}) and the explicit representations given in Eqs.~(\ref{epsandpres}) are typical for a momentum-space anisotropic system. The energy conservation equation (\ref{enmomcon1}) is satisfied if the energy densities calculated with the  non-equilibrium distribution functions $f$ or the equilibrium distribution function $f_{\rm eq}$ are equal, which requires that
\begin{eqnarray}
{\cal E}(\tau) &=& \frac{g_0}{\tau^2} \,
\int dP \, v^2\,  f(\tau,w,p_T) \, \nonumber \\
&=& \frac{g_0}{\tau^2}\,
\int dP \, v^2\,  f_{\rm eq}(\tau,w,p_T) \,
\nonumber \\
&=& 
 \frac{g_0 m^2 T }{\pi^2}  \left[ 3T K_{2}\left( \frac{m}{T} \right) +mK_{1} \left( \frac{m}{T}\right) \right]. \label{LM1}
\end{eqnarray}
This requirement represents the so-called dynamical Landau matching condition and can be used to define the effective temperature $T$ at any proper time. 

\section{Solutions of kinetic equation}
\label{sect:formsol}

We now proceed to solve the kinetic equation (\ref{kineq}) for a transversely homogenous boost-invariant system.

\subsection{General form of solutions}
\label{sect:genformsol}

The general form of solutions of Eq.~(\ref{kineq}) can be expressed as~\cite{Florkowski:2013lza,Florkowski:2013lya,
Baym:1984np,Baym:1985tna,Heiselberg:1995sh,Wong:1996va}
\begin{equation}
f(\tau,w,p_T) = D(\tau,\tau_0) f_0(w,p_T)  + \int_{\tau_0}^\tau \frac{d\tau^\prime}{\tau_{\rm eq}(\tau^\prime)} \, D(\tau,\tau^\prime) \, 
f_{\rm eq}(\tau^\prime,w,p_T) \, ,  \label{solG}
\end{equation}
where we have introduced the damping function
\begin{equation}
D(\tau_2,\tau_1) = \exp\left[-\int_{\tau_1}^{\tau_2}
\frac{d\tau^{\prime\prime}}{\tau_{\rm eq}(\tau^{\prime\prime})} \right].
\end{equation}
For the purposes of this paper, we will assume that at $\tau=\tau_0$ the distribution function $f$ can be expressed in Romatschke-Strickland form with an underlying Boltzmann distribution as the isotropic distribution~\cite{Romatschke:2003ms}
\begin{eqnarray}
f_0(w,p_T) &=& \frac{2}{(2\pi)^3}
\exp\left[
-\frac{\sqrt{(p\cdot u)^2 + \xi_0 (p\cdot z)^2}}{\Lambda_0} \, \right] \nonumber \\
&=& \frac{1}{4\pi^3}
\exp\left[
-\frac{\sqrt{(1+\xi_0) w^2 + (m^2+p_T^2) \tau_0^2}}{\Lambda_0 \tau_0}\, \right].
\label{G0}
\end{eqnarray}
This form simplifies to an isotropic Boltzmann distribution if the anisotropy parameter $\xi_0$ is zero, in which case the transverse momentum scale $\Lambda_0$ can be identified with the system's initial temperature $T_0$. 

\subsection{Dynamical Landau matching}
\label{sect:LM}

Multiplying Eqs.~(\ref{Geq}) and (\ref{G0}) by $g_0 v^2/\tau^2$ and integrating over momentum one obtains
\begin{eqnarray}
\frac{g_0}{\tau^2}\,\int dP \, v^2\,  f_{\rm eq}(\tau^\prime,w,p_T)
&=& \frac{g_0 T^4(\tau^\prime)}{2\pi^2} 
\, \tilde{\cal H}_2\left[ \frac{\tau^\prime}{\tau},\frac{m}{T(\tau^\prime)}\right],
\label{intGeq} \\
\frac{g_0}{\tau^2}\, \int dP \, v^2\,  f_0(w,p_T) 
&=& \frac{g_0 \Lambda^4_0}{2\pi^2} 
\, \tilde{\cal H}_2\left[ \frac{\tau_0}{\tau \sqrt{1+\xi_0}},\frac{m}{\Lambda_0}\right], \label{intG0}
\end{eqnarray}
where the function $\tilde{\cal H}_2(y,z)$ is defined by the integral
\begin{equation}
\tilde{\cal H}_2(y,z) = \int\limits_0^\infty du\, u^3 \, {\cal H}_2\left(y,\frac{z}{u} \right)
\, \exp\left(-\sqrt{u^2+z^2}\right) ,
\label{tildeH2}
\end{equation}
with
\begin{equation}
 {\cal H}_2(y,\zeta) = y \, 
 \int\limits_0^\pi d\phi \sin\phi
\, \sqrt{y^2 \cos^2\phi + \sin^2\phi + \zeta^2} \, . 
\label{H2}
\end{equation}

We note that Eqs.~(\ref{intGeq}) and (\ref{intG0}) are equal if $\tau = \tau^\prime = \tau_0$ and the system is initially isotropic ($\xi_0=0$). In this case the parameter $\Lambda_0$ can be identified with the system's temperature $T(\tau)$ and the expressions on the left-hand sides of (\ref{intGeq}) and (\ref{intG0}) become the equilibrium energy density ${\cal E}_{\rm eq}(T(\tau))$.  

In general, the integral appearing in Eq.~(\ref{H2}) can be performed analytically with the result being
\begin{equation}
{\cal H}_2(y,\zeta) 
= y \left( \sqrt{y^2+\zeta^2} + \frac{1+\zeta^2}{\sqrt{y^2-1}}
\tanh^{-1} \sqrt{\frac{y^2-1}{y^2+\zeta^2}} \, \right) ,
\label{H2an}
\end{equation}
however, the remaining integration over $u$ in Eq.~(\ref{tildeH2}) must be performed numerically.
We note that the function ${\cal H}_2(y,0)$ reduces to the function ${\cal H}(y)$ introduced in Ref.~\cite{Florkowski:2013lya}, and hence, $\tilde{\cal H}_2(y,0) = 6 {\cal H}(y)$.

Using Eqs.~(\ref{LM1}), (\ref{solG}), (\ref{intGeq}), and (\ref{intG0}) to implement the dynamical Landau matching, we obtain our main result
\begin{eqnarray}
&&  \hspace{-5mm}
2 m^2 T(\tau) \left[ 3 T(\tau) K_2\left(\frac{m}{T(\tau)}\right)
+ m K_1\left(\frac{m}{T(\tau)}\right) \right]
\nonumber \\
&& 
 = D(\tau,\tau_0) \Lambda^4_0 \tilde{\cal H}_2\left[ \frac{\tau_0}{\tau \sqrt{1+\xi_0}},\frac{m}{\Lambda_0}\right]  + \int\limits_{\tau_0}^\tau 
\frac{d\tau^\prime}{\tau_{\rm eq}} D(\tau,\tau^\prime)
T^4(\tau^\prime) 
\tilde{\cal H}_2\left[ \frac{\tau^\prime}{\tau},\frac{m}{T(\tau^\prime)}\right]. 
\label{LM2}
\end{eqnarray}
This is an integral equation for the effective temperature $T(\tau)$. It can be solved using the iterative method \cite{Banerjee:1989by}. In the massless limit ($m \to 0$), Eq.~(\ref{LM2}) reduces to Eq.~(38) of Ref.~\cite{Florkowski:2013lya}.

\subsection{Transverse and longitudinal pressures}
\label{sect:tandlpres}

A sensitive measure of the degree of equilibration can be obtained by computing the system's transverse and longitudinal pressures.  One can calculate the transverse and longitudinal pressures using Eqs.~(\ref{epsandpres}). Similarly to Eqs.~(\ref{intGeq}) and (\ref{intG0}) one obtains
\begin{eqnarray}
\frac{g_0}{2}\,\int dP \, p_T^2\,  f_{\rm eq}(\tau^\prime,w,p_T)
&=& \frac{g_0 T^4(\tau^\prime)}{4\pi^2} 
\, \tilde{\cal H}_{2T}\left[ \frac{\tau^\prime}{\tau},\frac{m}{T(\tau^\prime)}\right],
\label{intGeqT} \\
\frac{g_0}{2}\, \int dP \, p_T^2\,  f_0(w,p_T) 
&=& \frac{g_0 \Lambda^4_0}{4\pi^2} 
\, \tilde{\cal H}_{2T}\left[ \frac{\tau_0}{\tau \sqrt{1+\xi_0}},\frac{m}{\Lambda_0}\right], \label{intG0T}
\end{eqnarray}
where we have introduced the new function
\begin{equation}
\tilde{\cal H}_{2T}(y,z) = \int\limits_0^\infty du\, u^3 \, {\cal H}_{2T}
\left(y,\frac{z}{u} \right)
\, \exp\left(-\sqrt{u^2+z^2}\right) ,
\label{tildeH2T}
\end{equation}
with
\begin{eqnarray}
{\cal H}_{2T}(y,\zeta) &=&
y \, 
\int\limits_0^\pi \frac{d\phi \sin^3\phi }{
\, \sqrt{y^2 \cos^2\phi + \sin^2\phi + \zeta^2}} 
\label{H2T} \nonumber \\
\hspace{-1.75cm} &=& \frac{y}{(y^2-1)^{3/2}}
\left[\left(\zeta^2+2y^2-1\right) 
\tanh^{-1}\sqrt{\frac{y^2-1}{y^2+\zeta^2}}
-\sqrt{(y^2-1)(y^2+\zeta^2)} \right] . \hspace{5mm} 
\end{eqnarray}
Equations (\ref{intGeqT})--(\ref{H2T}) allow us to write a compact formula for the transverse pressure 
\begin{eqnarray}
{\cal P}_T(\tau) &=& \frac{g_0}{4\pi^2} D(\tau,\tau_0)
\Lambda_0^4 \tilde{\cal H}_{2T}\left[ \frac{\tau_0}{\tau \sqrt{1+\xi_0}},\frac{m}{\Lambda_0}\right]
\nonumber \\ && 
+ \frac{g_0 }{4 \pi^2} \int\limits_{\tau_0}^\tau 
\frac{d\tau^\prime}{ \tau_{\rm eq}} 
 D(\tau,\tau^\prime)
T^4(\tau^\prime) 
\tilde{\cal H}_{2T}\left[ \frac{\tau^\prime}{\tau},\frac{m}{T(\tau^\prime)}\right].
\label{PT}
\end{eqnarray}
In order to calculate ${\cal P}_T(\tau)$ using Eq.~(\ref{PT}), one has to determine the proper-time dependence of $T(\tau)$ by solving the integral equation (\ref{LM2}).  Once $T(\tau)$ is obtained, the integral over $\tau^\prime$ in (\ref{PT}) can be performed. Similarly to the functions $\tilde{\cal H}_2(y,z)$ and ${\cal H}_2(y,\zeta)$, the functions $\tilde{\cal H}_{2T}(y,z)$ and ${\cal H}_{2T}(y,\zeta)$ satisfy the relations ${\cal H}_{2T}(y,0)={\cal H}_{T}(y)$ and $\tilde{\cal H}_{2T}(y,0)=6 {\cal H}_{T}(y)$, where ${\cal H}_{T}(y)$ is defined in Ref.~\cite{Florkowski:2013lya}.

In the case of the longitudinal pressure, one can follow a similar procedure. 
Once again, one calculates the appropriate moments of the distribution functions
\begin{eqnarray}
\frac{g_0}{\tau^2}\,\int dP \, w^2\,  f_{\rm eq}(\tau^\prime,w,p_T)
&=& \frac{g_0 T^4(\tau^\prime)}{2\pi^2} 
\, \tilde{\cal H}_{2L}\left[ \frac{\tau^\prime}{\tau},\frac{m}{T(\tau^\prime)}\right],
\label{intGeqL} \\
\frac{g_0}{\tau^2}\, \int dP \, w^2\,  f_0(w,p_\perp) 
&=& \frac{g_0 \Lambda^4_0}{2\pi^2} 
\, \tilde{\cal H}_{2L}\left[ \frac{\tau_0}{\tau \sqrt{1+\xi_0}},\frac{m}{\Lambda_0}\right] ,
\label{intG0L}
\end{eqnarray}
where the function $\tilde{\cal H}_{2L}$ is defined by
\begin{equation}
\tilde{\cal H}_{2L}(y,z) = \int\limits_0^\infty du\, u^3 \, {\cal H}_{2L}
\left(y,\frac{z}{u} \right)
\, \exp\left(-\sqrt{u^2+z^2}\right) ,
\label{tildeH2L}
\end{equation}
with
\begin{eqnarray}
\hspace{-1.75cm} {\cal H}_{2L}(y,\zeta) &=& y^3 \, 
 \int\limits_0^\pi \frac{d\phi \sin\phi \cos^2\phi }{\, \sqrt{y^2 \cos^2\phi + \sin^2\phi + \zeta^2}}
\label{H2L} \\
\hspace{-1.75cm} &=& \frac{y^3}{(y^2-1)^{3/2}}
\left[
\sqrt{(y^2-1)(y^2+\zeta^2)}-(\zeta^2+1)
\tanh^{-1}\sqrt{\frac{y^2-1}{y^2+\zeta^2}} \,\,\right]. \nonumber
\end{eqnarray}
Using Eqs.~(\ref{intGeqL})--(\ref{H2L}) one finds
\begin{eqnarray}
{\cal P}_L(\tau) &=& \frac{g_0}{2\pi^2} D(\tau,\tau_0)
\Lambda_0^4 \tilde{\cal H}_{2L}\left[ \frac{\tau_0}{\tau \sqrt{1+\xi_0}},\frac{m}{\Lambda_0}\right]
\nonumber \\
&& + \frac{g_0 }{2 \pi^2} \int\limits_{\tau_0}^\tau 
\frac{d\tau^\prime}{ \tau_{\rm eq}}  D(\tau,\tau^\prime)
T^4(\tau^\prime) 
\tilde{\cal H}_{2L}\left[ \frac{\tau^\prime}{\tau},\frac{m}{T(\tau^\prime)}\right].
\label{PL}
\end{eqnarray}
Where, once again, if the function $T(\tau)$ is known, Eq.~(\ref{PL}) can be used to calculate the longitudinal pressure as a function of proper time. As before, one finds that ${\cal H}_{2L}(y,0)={\cal H}_{L}(y)$ and $\tilde{\cal H}_{2L}(y,0)=6 {\cal H}_{L}(y)$, where ${\cal H}_{L}(y)$ has been defined in \cite{Florkowski:2013lya}.

\section{Shear and bulk viscosities of a relativistic massive gas}
\label{sect:shearbulk}

In the results section we will compare the results of the exact solution with near-equilibrium expansions 
provided by first- and second-order viscous hydrodynamics.  In preparation for this, in this section we 
collect formulas for the shear and bulk viscosities of relativistic massive systems and discuss their 
asymptotic limits.

\subsection{Shear viscosity}
\label{sect:shear}

The shear viscosity of a classical massive gas in the relaxation time approximation (\ref{col-term}) was obtained originally by Anderson and Witting~\cite{Anderson:1974}
\begin{equation}
\eta(T) = \frac{\tau_{\rm eq} P_{\rm eq}(T)}{15} \, \gamma^3\left[ \frac{3}{\gamma^2}\frac{K_3}{K_2}  -\frac{1}{\gamma}+\frac{K_1}{K_2}-\frac{K_{i,1}}{K_2} \right] ,
\label{etaAW}
\end{equation}
where all functions above are understood to be evaluated at $\gamma \equiv m/T$ and the function $K_{i,1}$ is defined by the integral 
\begin{equation}
K_{i,1}(\gamma)= \int_0^{\infty} \frac{{\rm e}^{- \gamma \cosh t}}{\cosh t} \, dt \, ,
\label{Kin}
\end{equation}
which can be expressed as 
\begin{equation}
K_{i,1}(\gamma) = \frac{\pi}{2} \left[1 -
\gamma K_0(\gamma) L_{-1}(\gamma) - \gamma
K_1(\gamma) L_{0}(\gamma) \right] ,
\label{Kinan}
\end{equation}
where $L_i$ is a modified Struve function.
Equation (\ref{etaAW}) gives the proper-time dependence of the shear viscosity coefficient since, using the exact solution, one can determine $T(\tau)$.  This result will be compared with the kinetic estimate of the shear viscosity which can be obtained from
\begin{equation}
\eta_{\rm kin}(\tau) = \frac{1}{2} \, \tau \, \left({\cal P}_T(\tau)-{\cal P}_L(\tau)\right). 
\label{etakin}
\end{equation}
The form (\ref{etakin}) follows from the structure of the energy-momentum tensor in boost-invariant first-order viscous hydrodynamics. Therefore, one expects that the results obtained using Eqs.~(\ref{etaAW}) and (\ref{etakin}) will agree only at late times, $\tau \gg \tau_{\rm eq}$, when the system approaches equilibrium.

Since the temperature goes to zero at large times, proper understanding of the late-time asymptotic behavior of the system requires understanding of the $\gamma \rightarrow \infty$ limit of this quantity.
In this limit, Eq.~(\ref{etaAW}) becomes
\begin{equation}
\lim_{\gamma \rightarrow \infty} \eta = \tau_{\rm eq} {\cal P}_{\rm eq} + {\cal O}(\gamma^{-1}) \, ,
\label{etaAWlimit}
\end{equation}
where we have used the fact that
\begin{equation}
\lim_{\gamma \rightarrow \infty} \frac{K_{i,1}}{K_2} = 1 - \frac{5}{2\gamma} + \frac{39}{8\gamma^2} + \frac{45}{8\gamma^3} + \frac{885}{128\gamma^4} + {\cal O}(\gamma^{-5}) \, .
\end{equation}
As a consequence, in this limit $\bar{\eta} = \eta/{\cal S}_{\rm eq}$ becomes
\begin{equation}
\lim_{\gamma \rightarrow \infty} \bar{\eta }  
\approx \tau_{\rm eq} \,
\frac{{\cal P}_{\rm eq}}{{\cal S}_{\rm eq}}
\approx \tau_{\rm eq} \frac{T^2}{m} \, .
\label{etabarAWlimit}
\end{equation}

\subsection{Bulk viscosity}
\label{sect:bulk}

The bulk viscosity for a massive Boltzmann gas can be found in Refs.~\cite{Bozek:2009dw,Sasaki:2008fg,Romatschke:2011qp}
\footnote{Anderson and Witting \cite{Anderson:1974} also derived an expression for the bulk viscosity for a massive Boltzmann gas, however,
it does not match the result obtained by others.  In addition, their expression does not agree with our numerical results at late times, so we do not consider it here.}
\begin{equation}
\zeta(T) = \tau_{\rm eq} \, 
\frac{g_0 m^2}{3 \pi^2 T} 
\int_0^\infty p^2 e^{-\frac{\sqrt{m^2+p^2}}{T}}
\left[c_s^2(T) - \frac{p^2}{3(m^2+p^2)} \right]\,dp \, .
\label{zetaPBint}
\end{equation}
The integral over momentum in (\ref{zetaPBint}) can be performed giving
\begin{eqnarray}
\zeta(T) &=& \tau_{\rm eq} P_{\rm eq} \, 
\frac{\gamma^2}{3} \left[\left(c_s^2(T)-\frac{1}{3}\right) + \frac{\gamma}{3} \left(\frac{K_1}{K_2}-\frac{K_{i,1}}{K_2}\right) \right]
\nonumber \\
&=& \tau_{\rm eq} P_{\rm eq} \, 
\frac{\gamma^2}{3} 
\left[-
\frac{\gamma K_2}{3 (3 K_3+\gamma K_2)}
+ \frac{\gamma}{3} \left(\frac{K_1}{K_2}-\frac{K_{i,1}}{K_2}\right) \right] .
\label{zetaPB}
\end{eqnarray}
Since there are similar terms in the expressions for the shear and bulk viscosities (proportional to the difference $K_1-K_{i,1}$) one may find a relationship between $\zeta$ and $\eta$, namely
\begin{equation}
\zeta(T) = \frac{5}{3} \eta(T) - \tau_{\rm eq} P_{\rm eq} \, 
\frac{\gamma^3}{9} \left(\frac{K_2}{3 K_3+\gamma K_2}+\frac{3K_3}{\gamma^2 K_2} -\frac{1}{\gamma} \right).
\label{zetaeta1}
\end{equation}
In the limit of large masses (or, alternatively, low temperatures) one may use Eq.~(\ref{etaAWlimit}) and expand the Bessel functions on the right-hand side of Eq.~(\ref{zetaeta1}) to obtain
\begin{equation}
\lim_{\gamma \rightarrow \infty} \zeta(T) = \frac{2}{3} \tau_{\rm eq} P_{\rm eq} + {\cal O}(\gamma^{-1}) \, .
\end{equation}

Below, we present the numerical evidence that Eq.~(\ref{zetaPBint}) and its equivalent forms (\ref{zetaPB}) or (\ref{zetaeta1}) are the correct results for the bulk viscosity of a massive system. Our considerations are based on the analysis of the bulk viscous pressure $\Pi^{\rm kin}_\zeta$ which may be obtained directly from our exact solution by computing
\begin{equation}
\Pi^{\rm kin}_\zeta(\tau) = \frac{1}{3}
\left[{\cal P}_L(\tau) + 2 {\cal P}_T(\tau)
- 3 P_{\rm eq}(\tau) \right] .
\label{PIkz}
\end{equation}
This expression follows from the energy-momentum tensor used in boost-invariant viscous hydrodynamics  
and is not restricted to the first-order scheme.  Only when the system approaches equilibrium at proper times $\tau \gg \tau_{\rm eq}$, the bulk viscous pressure can be determined by the bulk viscosity $\zeta(T)$ through the relation 
\begin{equation}
\Pi^{\rm kin}_\zeta(\tau) \approx - \frac{\zeta(T(\tau))}{\tau} \, .
\label{1storderhyd}
\end{equation}

\subsection{Second-order viscous hydrodynamic equations for the bulk viscous pressure}
\label{sect:bulkhydro}

Our exact computation of the bulk viscous pressure can be compared with second-order viscous hydrodynamic predictions for the time dependence of this quantity.  Below we consider three possibilities for the evolution equation which appear in the literature:
\begin{equation}
\tau_{\rm eq} \frac{d\Pi^{\rm hyd}_\zeta}{d\tau} 
+ \Pi^{\rm hyd}_\zeta
=  - \frac{\zeta}{\tau} -\frac{\tau_{\rm eq} \Pi^{\rm hyd}_\zeta}{2} \left(\frac{1}{\tau}
- \frac{1}{\zeta} \frac{d\zeta}{d\tau} 
- \frac{1}{T} \frac{dT}{d\tau} \right), 
\label{adv}
\end{equation}
\begin{equation}
\tau_{\rm eq} \frac{d\Pi^{\rm hyd}_\zeta}{d\tau} 
+ \Pi^{\rm hyd}_\zeta
=  - \frac{\zeta}{\tau} -\frac{4 \tau_{\rm eq} \Pi^{\rm hyd}_\zeta}{3\tau}, 
\label{semadv}
\end{equation}
and
\begin{equation}
\tau_{\rm eq} \frac{d\Pi^{\rm hyd}_\zeta}{d\tau} 
+ \Pi^{\rm hyd}_\zeta
=  - \frac{\zeta}{\tau} \, .
\label{simple}
\end{equation}
These three forms appear in \cite{Muronga:2003ta,Heinz:2005bw}, \cite{Jaiswal:2013npa}, and \cite{Heinz:2005bw}, respectively.
The final expression (\ref{simple}) is an approximation to the first expression (\ref{adv}) which is obtained by discarding the second term on the right hand side.  In the subsequent results section we numerically
solve Eqs.~(\ref{adv})-(\ref{simple}) using the proper-time dependence of the effective temperature $T(\tau)$ obtained from the exact solution and then compare to the bulk pressure extracted directly from the exact solution using Eq.~(\ref{PIkz}).\footnote{We have checked explicitly that using the proper-time dependence of the temperature from second-order viscous hydrodynamics yields the same result for the bulk pressure to within a fraction of a percent for the values of $\tau_{\rm eq}$ used herein.}

\section{Results}
\label{sect:results}

In this section we present results of our exact solution for a specific initial condition and set
of physical parameters.  We compare the massless and massive exact solutions to determine what effect the mass has on the evolution of the system.  We then compare the shear and bulk viscosities from the literature with those extracted from the exact solution by considering the late-time near-equilibrium evolution of the solutions.  Finally, we compare the evolution of the bulk pressure from the exact solution with the evolution predicted by three different viscous hydrodynamics approaches.

\subsection{Initial conditions}
\label{sect:initcond}

We perform our numerical calculations for two fixed values of the initial effective temperature: $T_0$ = 600 MeV and $T_0$ = 300 MeV. The equilibration time $\tau_{\rm eq}$ is kept constant and equal to 0.5 fm/c.  The integral equation (\ref{LM2}) is solved by the iterative method.  The initial time is taken to be \mbox{$\tau_0$ = 0.5 fm/c} and we continue the evolution until \mbox{$\tau$ = 10 fm/c}. In order to identify the mass effects more clearly, we consider the case of a fixed mass with \mbox{$m$ = 300 MeV}.\footnote{In the context of quasiparticle models which assume a gluon mass $m_g \sim g T$, with $g \sim 2$ at phenomenologically relevant temperatures, such a mass might even be a bit small.}  The degeneracy factor $g_0$ is taken to be 16, however, the specific value of $g_0$ is irrelevant for our conclusions since it either cancels in ratios we consider or appears as an overall scaling.

The initial distribution function is assumed to be of Romatschke-Strickland form \cite{Romatschke:2003ms} with the initial anisotropy parameter $\xi_0 \in \{0,100\}$, corresponding to an initially isotropic or oblate initial configuration, respectively. The transverse-momentum scale $\Lambda_0$ is chosen in such a way that the initial energy density of an anisotropic system coincides with the energy density of an equilibrium system with temperature $T_0$
\begin{equation}
2 m^2 T_0 
 \left[ 3T_0 K_{2}\!\left( \frac{m}{T_0} \right) +m K_{1}\!\left( \frac{m}{T_0}\right) \right]
 =  \Lambda^4_0
\, \tilde{\cal H}_2\left[ \frac{1}{ \sqrt{1+\xi_0}},\frac{m}{\Lambda_0}\right],
\label{initL0}
\end{equation} 
which is simply the Landau matching condition (\ref{LM2}) at $\tau=\tau_0$. We note that for fixed $T_0$ and $\xi_0$ the value of $\Lambda_0$ depends on $m$. In the special case $m=0$ Eq.~(\ref{initL0}) reduces to the form
\begin{equation}
2 T_0^4 = \Lambda_0^4 \, {\cal H}\!\left(
\frac{1}{\sqrt{1+\xi_0}}
\right) ,
\label{initL00}
\end{equation}
where, as mentioned previously, ${\cal H}$ is defined in Ref.~\cite{Florkowski:2013lya}.

\begin{figure}[t]
\centerline{\includegraphics[angle=0,width=0.7\textwidth]{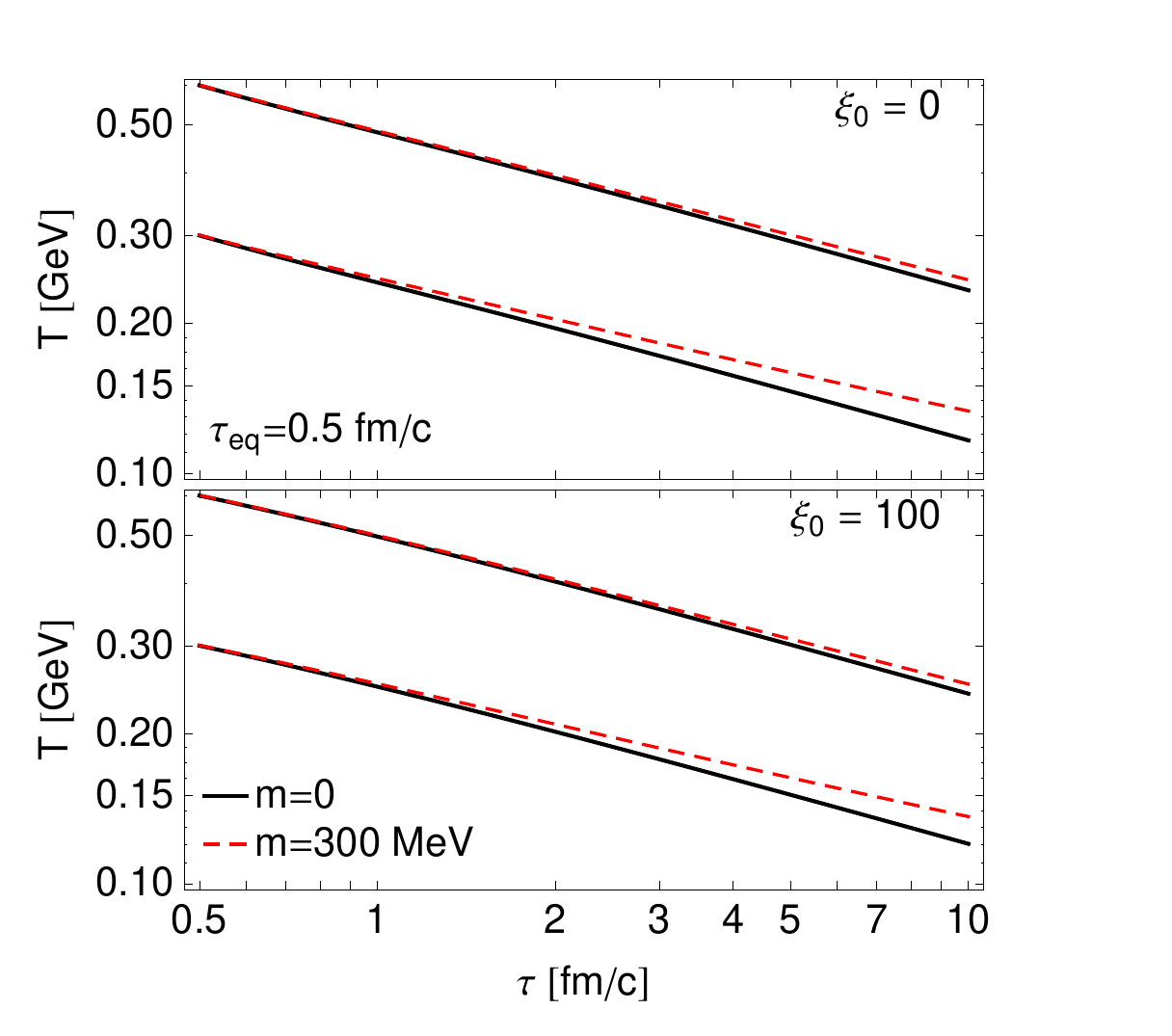}}
\caption{(Color online) Time dependence of the effective temperature $T$.  Solid lines are the solution for $m=0$ and dashed lines are the solution for $m = 300$ MeV.  In both the top and bottom panels, the upper set of curves corresponds to \mbox{$T_0 = 600$ MeV}, while the bottom set of curves corresponds to $T_0 = 300$ MeV.}
\label{fig:Temp}
\end{figure}

\subsection{Effective temperature}
\label{sect:efftemp}

In Fig.~\ref{fig:Temp} we plot the time dependence of the effective temperature $T$ obtained by iterative solution of Eq.~(\ref{LM2}).  In the top panel we show the results obtained for an initially isotropic system and in the bottom panel we show the case of a highly oblate initial anisotropy.  In both the top and bottom panels, the solid lines are the solution for $m=0$ and the dashed lines are the solution for $m = 300$ MeV.  Also, in both the top and bottom panels, the upper set of curves corresponds to $T_0 = 600$ MeV, while the bottom set of curves correspond to $T_0 = 300$ MeV.  As we can see from this figure, the primary effect of the mass on the effective temperature is to cause it decrease more slowly as a function of proper time.  This behavior is consistent with what one expects from hydrodynamics since, as the mass increases, the speed of sound decreases causing the energy density (and hence the effective temperature) to decrease more slowly as a function of proper time.  We also note that the effect of adding a mass is larger for lower initial temperatures, as one would expect based on general arguments.  

\begin{figure}[t]
\centerline{\includegraphics[angle=0,width=0.7\textwidth]{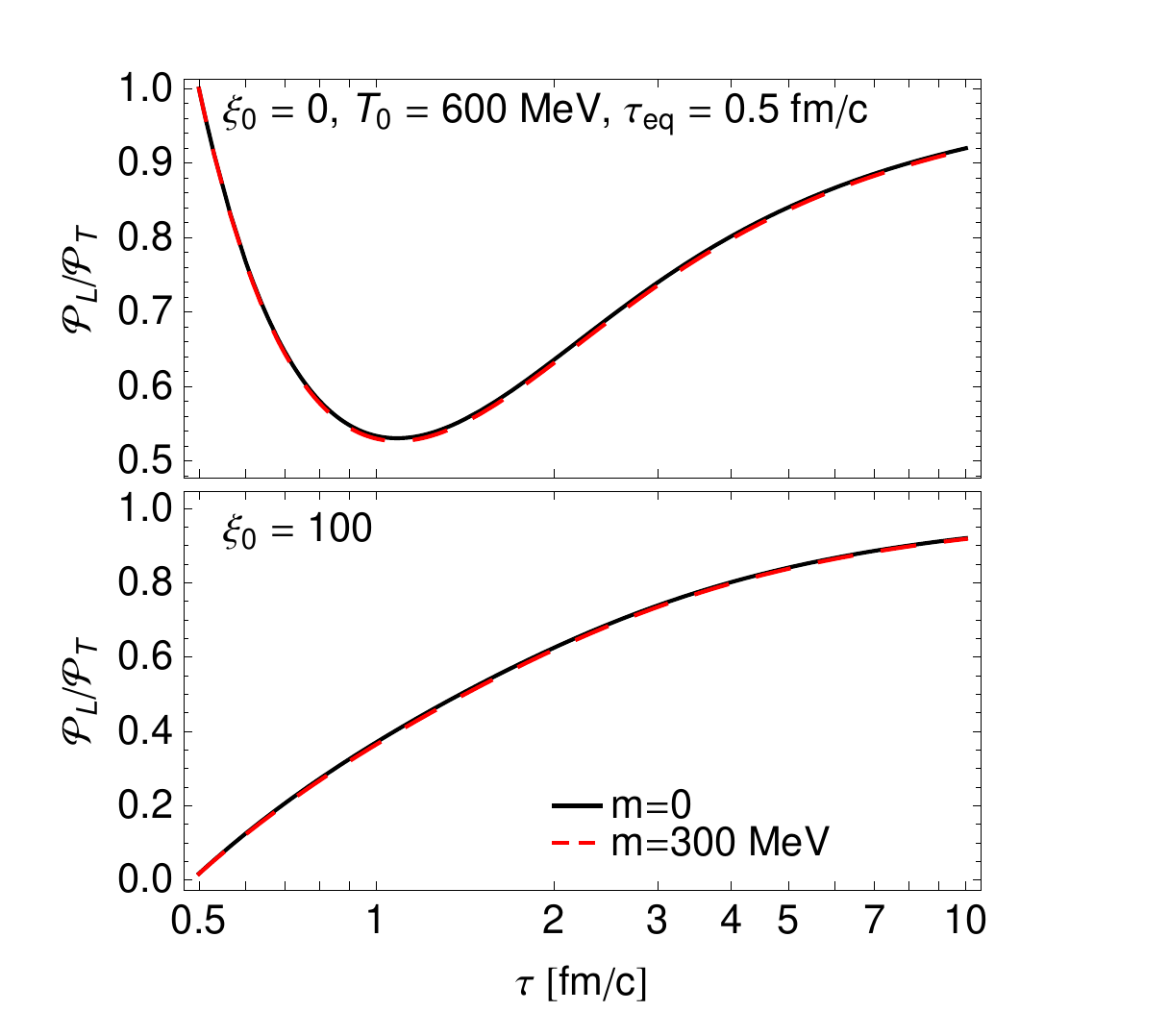}}
\caption{(Color online) The time dependence of the ratio of the longitudinal and transverse pressures.
The initial temperature was taken to be $T_0 = 600$ MeV.  The top panel shows the case of an initially isotropic system and the bottom panel shows the case of an initially oblate system.}
\label{fig:PL2PT}
\end{figure}

\subsection{Pressure Anisotropy}
\label{sect:pt2pl}

In Fig.~\ref{fig:PL2PT} we plot the time dependence of the ratio of the longitudinal and transverse 
pressures ${\cal P}_L/{\cal P}_T$ obtained by using the iterative solution of Eq.~(\ref{LM2}) to 
evaluate Eqs.~(\ref{PT}) and (\ref{PL}).  The initial temperature was taken to be $T_0 = 600$ MeV, however, we note that this ratio depends very weakly on the initial temperature when the relaxation time is a constant.  In the top panel of Fig.~\ref{fig:PL2PT} we show the case 
of an initially isotropic system and in the bottom panel we show the case of an initially oblate system.  
As before, the solid lines are the solution for $m=0$ and the dashed lines are the solution for 
$m = 300$ MeV.  As we can see from these figures, having a non-zero mass seems to have very little 
effect on the pressure anisotropy.

\begin{figure}[t]
\centerline{\includegraphics[angle=0,width=0.7\textwidth]{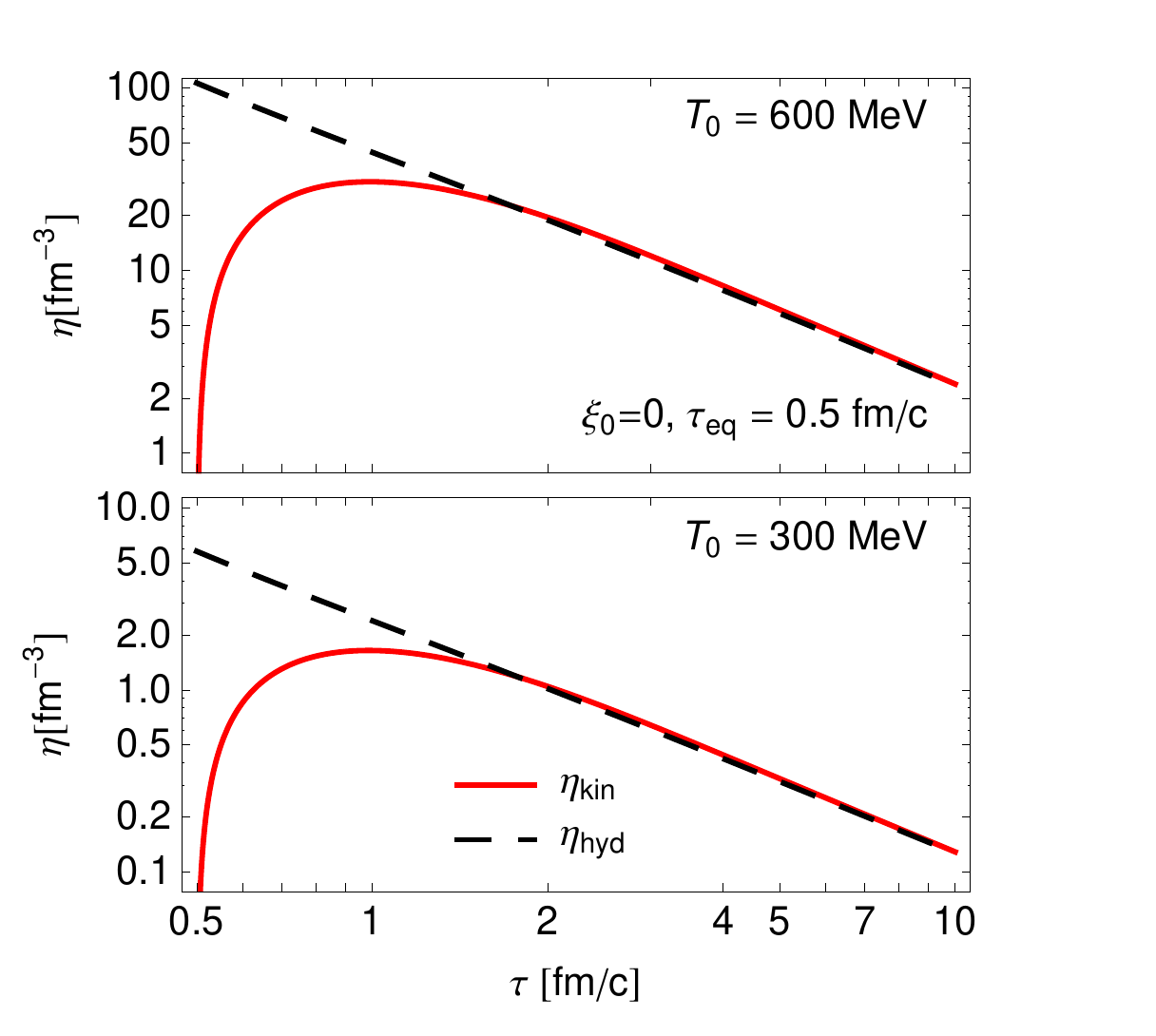}}
\caption{(Color online) 
The effective shear viscosity $\eta_{\rm kin}$ (solid line) compared with $\eta_{\rm hyd}$ 
(dashed line) obtained from Eq.~(\ref{etaAW}).  The system was assumed to be initially isotropic, 
i.e. $\xi_0=0$.  The top panel shows the results obtained for $T_0 = 600$ MeV and the bottom panel 
shows the results obtained for $T_0 = 300$ MeV.
}
\label{fig:shear_xi0}
\end{figure}
\begin{figure}[t]
\centerline{\includegraphics[angle=0,width=0.7\textwidth]{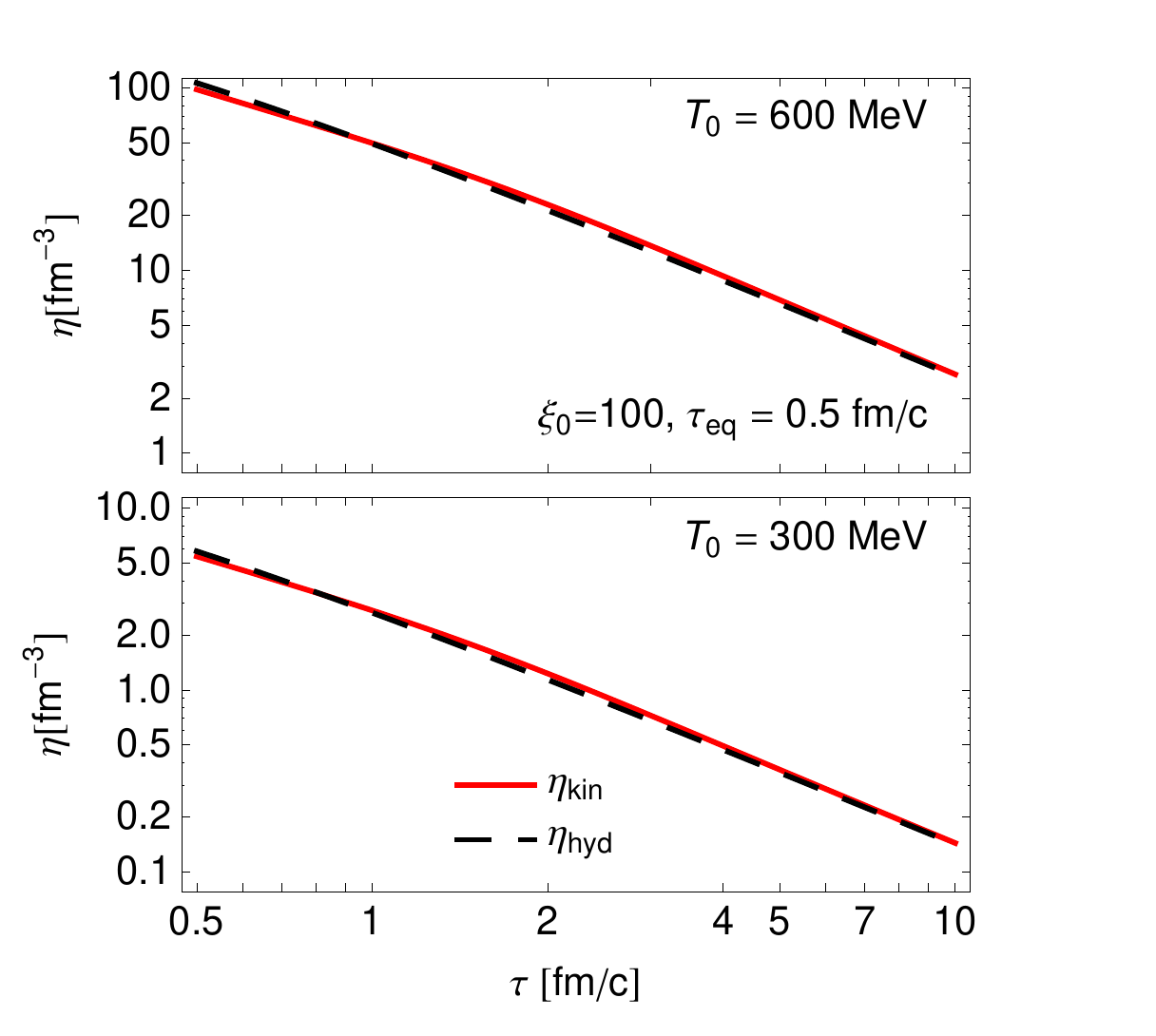}}
\caption{(Color online) 
Same as Fig.~\ref{fig:shear_xi0} except with $\xi_0 = 100$.
}
\label{fig:shear_xi100}
\end{figure}

\subsection{Shear viscosity}
\label{sect:shearres}

We now turn to a comparison of the effective shear viscosity extracted from our exact solution using
Eq.~(\ref{etakin}) with the near-equilibrium behavior predicted by viscous hydrodynamics.  In
Figs.~\ref{fig:shear_xi0} and \ref{fig:shear_xi100} we plot the resulting $\eta_{\rm kin}$ 
compared with $\eta_{\rm hyd}$ obtained from Eq.~(\ref{etaAW}).  Figure~\ref{fig:shear_xi0} shows the case $\xi_0=0$ and Fig.~\ref{fig:shear_xi100} shows the case $\xi_0=100$.  In both
figures the top panel shows the results obtained for $T_0 = 600$ MeV and the bottom panel shows the 
results obtained for $T_0 = 300$ MeV.  As we can see from these figures, after some initial transient non-equilbrium evolution during which the effective shear viscosity deviates from the near-equilibrium value, the results converge and the exact solution is well-approximated by the near-equilibrium shear viscosity (\ref{etaAW}).  

\subsection{Bulk viscosity and pressure}
\label{sect:bulkres}

\begin{figure}[t]
\centerline{\includegraphics[angle=0,width=0.7\textwidth]{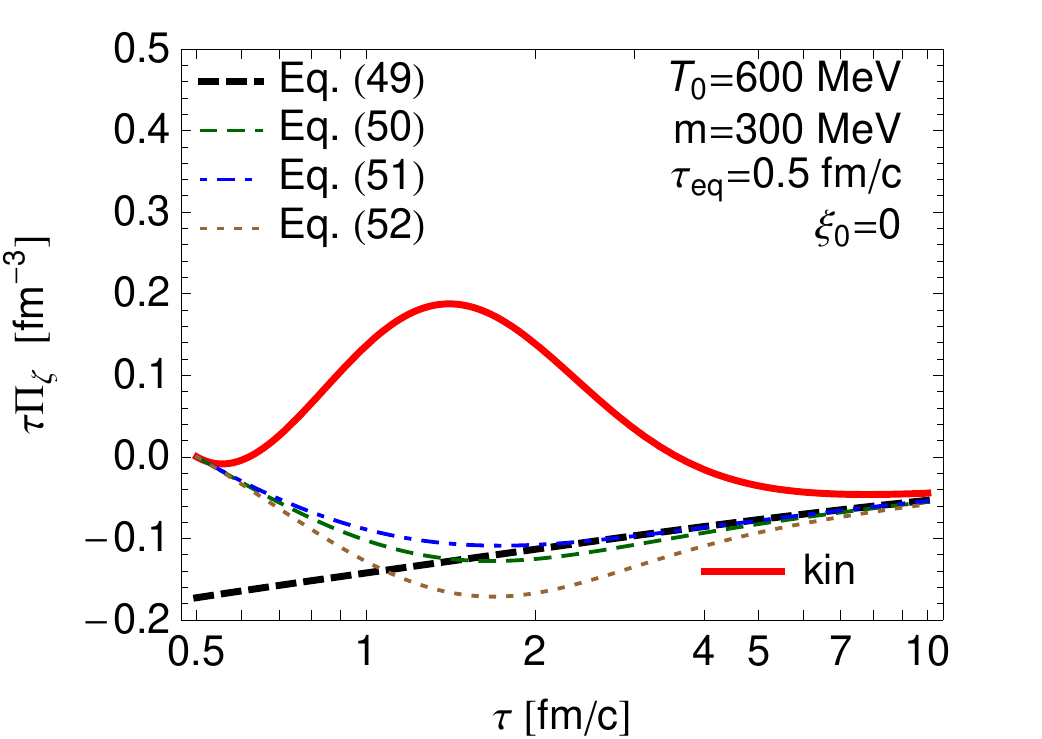}}
\caption{(Color online) Proper-time dependence of the bulk pressure times $\tau$
for $\xi_0=0$ and \mbox{$T_0=600$ MeV}.  Solid line is the exact solution obtained from Eq.~(\ref{PIkz}).  The other curves correspond to the first-order solution (\ref{1storderhyd}) and the solutions
of Eqs.~(\ref{adv}), (\ref{semadv}), and (\ref{simple}).
}
\label{fig:bulk_T600_xi0}
\end{figure}

\begin{figure}[t]
\centerline{\includegraphics[angle=0,width=0.7\textwidth]{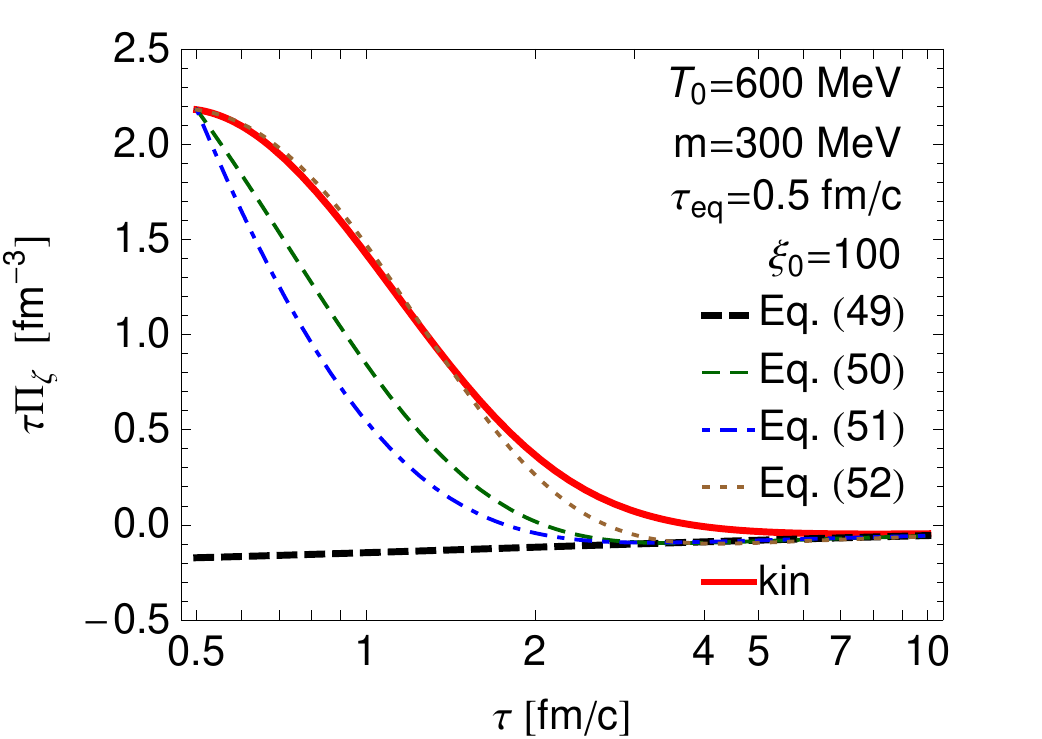}}
\caption{(Color online) Same as Fig.~\ref{fig:bulk_T600_xi0} except with $\xi_0=100$.}
\label{fig:bulk_T600_xi100}
\end{figure}

\begin{figure}[t]
\centerline{\includegraphics[angle=0,width=0.7\textwidth]{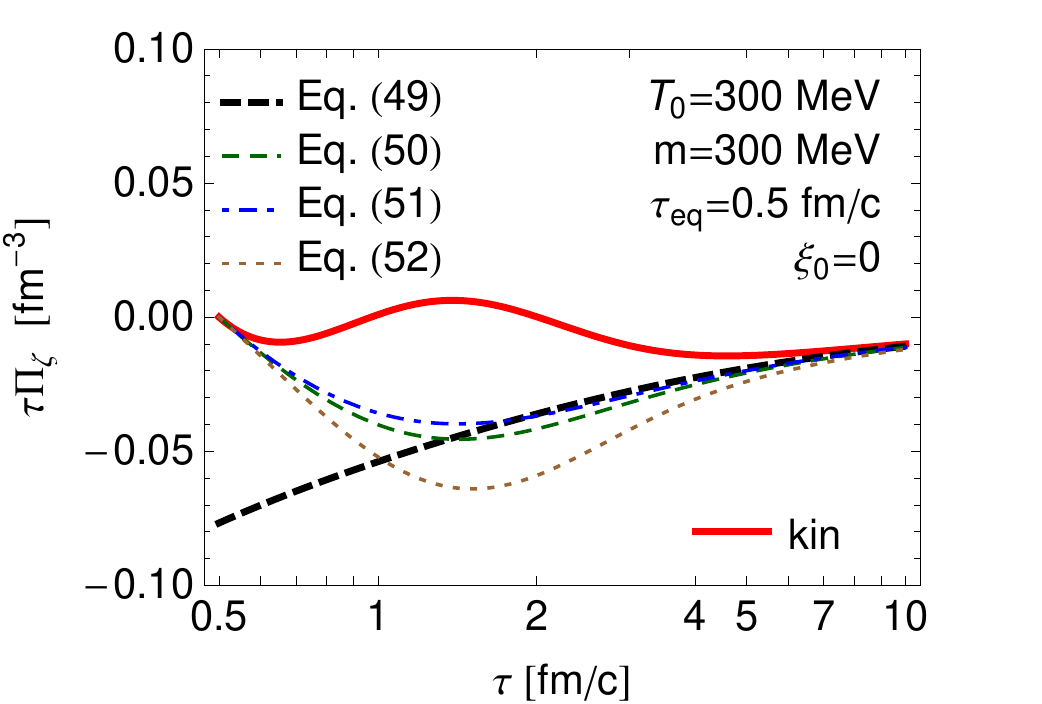}}
\caption{(Color online) Same as Fig.~\ref{fig:bulk_T600_xi0} except with $T_0=300$ MeV.}
\label{fig:bulk_T300_xi0}
\end{figure}

\begin{figure}[t]
\centerline{\includegraphics[angle=0,width=0.7\textwidth]{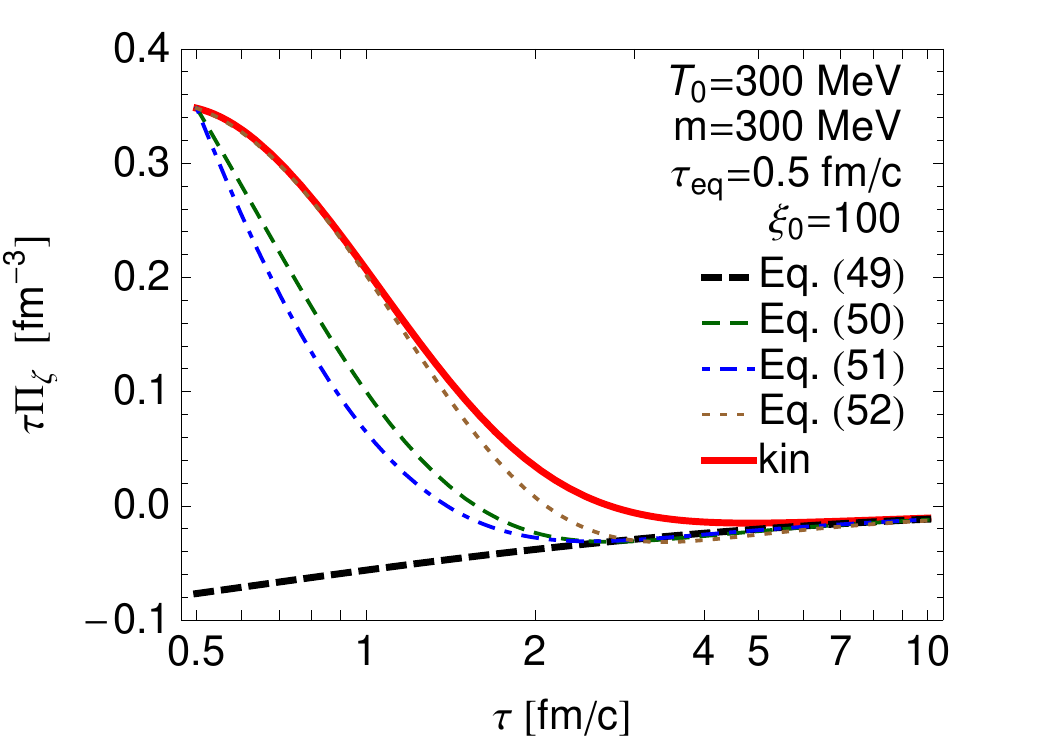}}
\caption{(Color online) Same as Fig.~\ref{fig:bulk_T600_xi0} except with $T_0=300$ MeV and $\xi_0=100$.}
\label{fig:bulk_T300_xi100}
\end{figure}

We now turn to comparison of the proper-time dependence of the bulk pressure and associated bulk
viscosity extracted from our exact solution using Eq.~(\ref{PIkz}).  In 
Figs.~\ref{fig:bulk_T600_xi0}--\ref{fig:bulk_T300_xi100} we plot the bulk pressure times $\tau$
for five different cases.  The solid line is the result obtained using the exact solution and 
Eq.~(\ref{PIkz}).  The other curves shown correspond to:  the first-order solution (\ref{1storderhyd})
indicated by a thick dashed line and the solutions to Eqs.~(\ref{adv}), (\ref{semadv}), and 
(\ref{simple}) indicated by a thin dashed line, a dot-dashed line, and a dotted line,
respectively.  As we can see from these figures, the exact solution and all second-order viscous 
hydrodynamics variations tend toward the first-order solution at late times.  However, none of the
second-order viscous hydrodynamics variations seems to accurately describe the early-time evolution
of the bulk viscous pressure in all cases.  Paradoxically, the simple approximate form (\ref{simple}) 
seems to provide the best approximation when the system initially possesses a highly oblate momentum-space
anisotropy, however, it provides the worst approximation if the system is initially isotropic in
momentum space.  These results indicate that there may be something incomplete in the manner in which
second-order viscous hydrodynamics treats the bulk pressure.  One possibility is that the evolution
equations for the bulk pressure used herein have neglected to include the possibility of shear-bulk
coupling which appears, for example, in the complete expansion derived in Ref.~\cite{Denicol:2012cn}.


\section{Temperature-dependent relaxation time}
\label{sect:taeqT}

Before concluding, we would like to point out that in the previous section we considered numerical
results obtained using a time-independent relaxation time $\tau_{\rm eq}$, however, our exact 
solution (\ref{LM2}) is not limited to this case.  If one wanted to study the case, for example,
that the ratio of the shear viscosity to entropy density were held fixed, this would imply a 
temperature-dependent, and hence time-dependent, relaxation time.  For general masses one could
use Eq.~(\ref{etaAW}) expressed in terms of $\bar\eta = \eta/S_{\rm eq}$ and then solve for 
$\tau_{\rm eq}$ as function of the mass and temperature.  
If this is done, one would find that 
the relaxation time depends non-trivially on the assumed mass.  The relation becomes particularly
transparent in the limit of large masses, in which case one can use the asymptotic form 
(\ref{etabarAWlimit}) to obtain $\lim_{\gamma \rightarrow \infty} \tau_{\rm eq} = m \bar\eta/T^2$,
which implies that, for fixed temperature, the relaxation time goes to infinity.
In practice, this means that for a massive system one will see larger deviations from equilibrium 
than for a massless system if one fixes $\bar\eta$ and compares the two.

\section{Conclusions}
\label{sect:concl}

In this paper we generalized the results of Refs.~\cite{Florkowski:2013lza,Florkowski:2013lya} 
to a system of massive particles obeying Boltzmann statistics.  Our main result is an integral 
equation (\ref{LM2}) that can be solved to arbitrary numerical precision using the method of 
iteration.  Based on this solution one can obtain the proper-time dependence of the full 
one-particle distribution function and, as a consequence, one can numerically obtain all 
thermodynamic functions to arbitrary numerical precision.  We presented explicit expressions for 
the transverse pressure (\ref{PT}) and longitudinal pressure (\ref{PL}).  We then presented
the results of numerical solution of the integral equation for the effective temperature,
the pressure anisotropy, the effective shear viscosity, and the bulk pressure.  We found that
the effect of finite masses on the effective temperature is to cause it to decrease more slowly
in proper time, which is consistent with hydrodynamic expectations.  We found that the pressure 
anisotropy depends very weakly on the mass in the case that $\tau_{\rm eq}$ is assumed to be 
independent of the temperature.  Finally, we compared our exact results with results obtained 
from relativistic hydrodynamics.  We found that the standard expressions available in the 
literature for the mass and temperature dependence of the shear and bulk viscosities correctly 
describe the evolution of the system well for $\tau \gg \tau_{\rm eq}$.  

Looking forward it will be interesting to compare results obtained using anisotropic 
hydrodynamics for the massive case with the exact solution obtained herein.  There are now two 
formulations of leading-order anisotropic hydrodynamics on the market; one which uses the zeroth
moment of the Boltzmann equation to obtain an equation of motion \cite{Martinez:2010sc}
and one which uses the second moment of the Boltzmann equation to obtain an equation of motion
\cite{Tinti:2013vba}.  The exact solution continued here can be used to determine which scheme
provides the best approximation.  We leave this for future work.

\acknowledgments

We thank M. Martinez for discussions.
R.R. was supported by Polish National Science Center grant No. DEC-2012/07/D/ST2/02125, the 
Foundation for Polish Science, and U.S.~DOE Grant No. DE-SC0004104.   W.F. and E.M. were 
supported by Polish National Science Center grant No. DEC-2012/06/A/ST2/00390.  M.S. was 
supported in part by U.S.~DOE Grant No. DE-SC0004104.

\bibliography{rtamass}

\end{document}